\documentclass[a4paper,10pt]{article}

	\usepackage{enumerate,enumitem,hyperref,siunitx}
	\usepackage{bm,amsmath,amssymb,amsthm,physics}
	\usepackage{float,authblk,setspace,lineno}
	\usepackage{cite,graphicx}
	\usepackage[symbol]{footmisc}
    \usepackage{combelow}
	\usepackage[font=small,labelfont=bf,figurename=Figure,labelsep=period]{caption}
	\usepackage[margin=1in]{geometry}
	\usepackage[capitalize]{cleveref}

	\newenvironment{keywords}
    {\footnotesize \textbf{Keywords:} }
    { \vspace{0.2cm} }

\begin{document}

	\title{A cautionary tale of model misspecification and identifiability}


	\author[1*]{Alexander P Browning}
	\author[1]{Jennifer A Flegg}
	\author[2*]{Ryan J Murphy}
	\affil[1]{School of Mathematics and Statistics, University of Melbourne, Australia}
	\affil[2]{UniSA STEM, The University of South Australia, Mawson Lakes, SA 5095, Australia}
	

\date{\today}
\maketitle
\footnotetext[1]{Corresponding authors: apbrowning@unimelb.edu.au and ryan.murphy@unisa.edu.au.}


	\begin{abstract}
		\noindent Mathematical models are routinely applied to interpret biological data, with common goals that include both prediction and parameter estimation. A challenge in mathematical biology, in particular, is that models are often complex and non-identifiable, while data are limited. Rectifying identifiability through simplification can seemingly yield more precise parameter estimates, albeit, as we explore in this perspective, at the potentially catastrophic cost of introducing model misspecification and poor accuracy. We demonstrate how uncertainty in model structure can be propagated through to uncertainty in parameter estimates using a semi-parametric Gaussian process approach that delineates parameters of interest from uncertainty in model terms. Specifically, we study generalised logistic growth with an unknown crowding function, and a spatially resolved process described by a partial differential equation with a time-dependent diffusivity parameter. Allowing for structural model uncertainty yields more robust and accurate parameter estimates, and a better quantification of remaining uncertainty. We conclude our perspective by discussing the connections between identifiability and model misspecification, and alternative approaches to dealing with model misspecification in mathematical biology.
	\end{abstract}


\section{Introduction}

Mathematical models are now routine in the interpretation of biological data. Model parameters, estimated through calibration, provide a quantitative characterisation of behaviour that cannot be directly measured \cite{Liepe.2014,Gabor.2015}: for example, per-capita growth or movement rates in the presence of crowding from population-level data \cite{Jin.2016a,Simpson.2022}. Comparison of parameter estimates between experimental conditions, therapeutic treatments, or species allows practitioners to statistically quantify physiological change. Moreover, the resulting calibrated models can be used for prediction, to validate model assumptions, and in experimental design \cite{Faller.2003,Mogilner.2006}.

In the context of experimental data, it is often assumed requisite that models be identifiable and that parameter estimates can be precisely determined. Both structural and practical identifiability analysis are now widely accepted, rightly, as a part of the model development and eventual model calibration workflow \cite{Raue.2009,Miao.2011}. Resolution of non-identifiability, however, remains a challenge. Typical suggestions in the literature are model reduction, reparameterisation, or simplification \cite{Raue.2009}; to increase the quantity or quality of data \cite{Porthiyas.2024}; or to modify the type of measurements that can be observed (for example, to additionally observe spatial information or previously unobserved species) \cite{Walter.1981}. A challenge in mathematical biology, however, is that available data are often collected before any analysis, or otherwise that experimental constraints limit the type of measurements that can be observed. Practitioners who wish to resolve identifiability issues are often, therefore, left with little choice but to resort to model simplification.

While resolving identifiability and potentially improving the precision of parameter estimates, model simplification can exacerbate model misspecification \cite{Dennis.2019}. To visualise the effect of model misspecification on parameter inference, we first consider a process characterised by the generalised logistic growth model for cell density $u(t)$
	\begin{equation}\label{richards}
		\dv{u}{t} = r u f\!\left(\dfrac{u}{K}\right),\qquad f(\hat{u}) = \left(1 - \hat{u}^\beta\right),\qquad u(0) = u_0,
	\end{equation}
	where $f(\hat{u})$ is the \textit{crowding function} and $K$ is the carrying capacity. The specific form of $f(\hat{u})$ in \cref{richards} is commonly referred to as Richards model \cite{Richards.1959}. In our previous work \cite{Simpson.2022}, we demonstrate that calibration of the full parameter set $(r,K,\beta)$ to data can lead to issues with practical identifiability, manifesting as a strong dependence between the exponent parameter $\beta$ and the low-density growth rate parameter $r$. The specific choice of $\beta$ is often based on convention, and not data analysis: in cell and cancer biology, the canonical logistic model ($\beta = 1$) and to a lesser extent the Gompertz model $(\beta \rightarrow 0)$ are routine \cite{Maini.2004,Gerlee.2013,Sarapata.2014}. The von Bertalanffy model is historically associated with fish growth, although selection between various parametric models is now accepted practice in this field \cite{Katsanevakis.2008}. Although not always the result of an identifiability analysis, restricting study to a simplified (e.g. logistic) model can yield improved parameter identifiability and estimate precision, however the introduced misspecification is potentially problematic for parameter estimation accuracy \cite{White.1982}.

\begin{figure}[!t]
	\centering
	\includegraphics{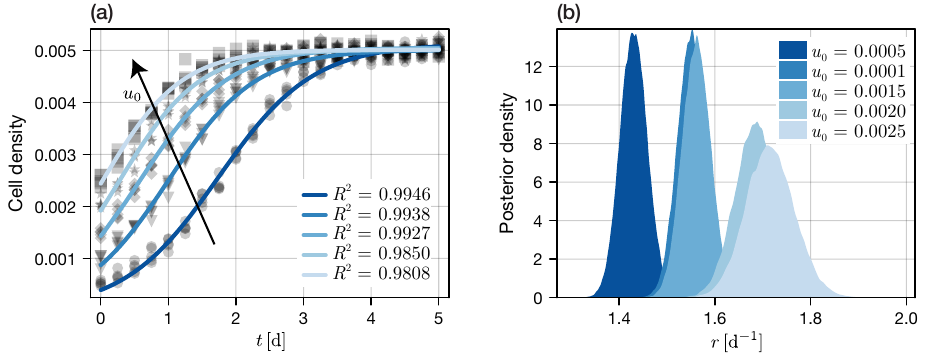}
	\caption[Figure 1]{Unreliable parameter estimates drawn from an identifiable and misspecified model. Synthetic data generated from Richards model (\cref{richards} with $\beta = 2$) with a range of initial conditions $u(0) = u_0$ is calibrated to the misspecified canonical logistic model (\cref{richards} with $\beta = 1$). (a) Synthetic data is produced at the true parameter set $(r,K,u_0,\sigma) = (1,\SI{5e-3},u_0,10^{-4})$ for various $u_0$ (see legend) using Richards model subject to additive Gaussian noise with variance $\sigma^2$. The parameter set $(r,K,u_0,\sigma)$ is calibrated to the canonical logistic using a Bayesian inference scheme, with model fit at the maximum a posteriori estimate for each initial condition shown. Goodness-of-fit is quantified using the expected value of the Bayesian $R^2$ statistic \cite{Gelman.2019}. (b) Due to model misspecification, the marginal posterior distributions for the low-density proliferation rate $r$ depend strongly on the initial condition, yet the parameter is practically identifiable in all cases. }
	\label{fig1}
\end{figure}
    
	To continue our demonstration, we consider data collected from a routinely performed \textit{cell proliferation assay}, where cells are placed in a dish and allowed to grow to carrying capacity. This experiment is often performed to estimate the growth rate of a population of cells. Analogous data are routinely collected throughout biology and ecology \cite{Simpson.2022}. In \cref{fig1}a, we generate synthetic cell density data from Richards model with $\beta = 2$ for a range of initial cell densities. We proceed to calibrate the identifiable logistic model (i.e., misspecified, with $\beta = 1$) to the generated data using a standard Bayesian inference approach that we later describe. The resultant model fits shown in \cref{fig1}a are quantitatively excellent (as measured using the expected Bayesian $R^2$ statistic \cite{Gelman.2019}), with the parameter $r$ being practically identifiable for all data sets (\cref{fig1}b). However, misspecification results in a strong dependence between estimates of $r$ and the initial condition, $u_0$. Consequently, any statistical analysis between experiments initiated with different cell densities would incorrectly suggest a physiological difference between cell populations, despite the underlying behaviour being identical. Parameter estimates might be considered precise, but they are not accurate. 

	Issues relating to misspecification can be detected from residual analysis \cite{Simpson.2022,Lambert.2023} or, should a sufficient quantity of data be available, through model validation. Fundamentally, however, issues relating to misspecification can arise from the inability of a (potentially) simple and identifiable model to capture the true underlying dynamics \cite{Amadi.2023}. Even relaxing assumptions on $\beta$ and calibrating Richards model to cell growth data is potentially problematic: both in terms of the prior that must be placed on the purely phenomenological exponent $\beta$ and the assumed parametric form for the crowding function. Significant effort in recent years has focused on relaxing assumptions related to the functional form of terms in differential equations \cite{Metzcar.2024}. Symbolic regression, a parametric approach, attempts to ``learn'' the functional form of an ordinary or partial differential equation using a library of potential terms, applying sparse regression to avoid overfitting \cite{Brunton.2016}.  Non-parametric approaches, meanwhile, include those that represent one or more terms using a Gaussian process or a neural network \cite{Wang.2014,Heinonen.2018,Chen.2019,Yang.2021,Dandekar.2022,Bhouri.2022,VandenHeuvel.2022}, or through expansion or spline representation of any unknown functions \cite{Cao.2008,Cao.2012,Paul.2011, Chen.2017}. Recently introduced in the context of differential equation modelling in biology, Biological Informed Neural Networks (BINNs) \cite{Lagergren.2020} constrain neural networks through a biologically informed model structure; for example, a reaction-diffusion equation with density-dependent diffusivity and reaction rate. All the aforementioned approaches alleviate model misspecification; however, they are often heavily parameterised and data intensive. With a modest amount of experimental data, it remains unclear how these functional-form-free approaches affect the identifiability of key quantities of interest. 

	In this perspective, we employ discretised Gaussian processes as a straightforward non-parametric approach to avoid model misspecification and allow for uncertainty in model structure. Gaussian processes have a long history of being applied to address model misspecification in the analysis of physical systems \cite{Kennedy.2001,Brynjarsdottir.2014}. Many other approaches allow practitioners to avoid specifying (potentially incorrect) functional parametric forms \cite{Cao.2008,Paul.2011,Cao.2012,Chen.2017,Chen.2019,Dandekar.2022}. We base our perspective on a Gaussian process approach as it provides both a natural way of incorporating and visualising prior knowledge, and allows us to directly quantify the uncertainty in inferred parameters and functions. Specifically, we replace specific terms in a differential equation with a function on which we can place a clear prior: for example, our prior knowledge may be that candidate crowding functions are, on average, logistic. Importantly, we retain the ability to perform model calibration using standard Bayesian approaches, such that we can compare parameter identifiability with that from a potentially misspecified model. We aim in this perspective to demonstrate the effect of prior knowledge and misspecification on parameter estimation. Our goal is to accurately estimate the \textit{uncertainty} in parameter estimates, rather than to obtain precise and identifiable---but potentially biased---estimates of model parameters.

	We demonstrate and discuss our approach on both a temporal and a spatio-temporal model. First, we apply our approach to estimate low-density growth rates from cell density data using a non-parametric generalised logistic model, for which we possess an analytical solution. Next, we repeat this analysis using synthetic cell density data collected from a spatially heterogeneous process: in this case, we are guaranteed that our parametric models are misspecified. Finally, we employ our approach to account for an unknown, time-dependent diffusivity in the estimation of the low-density growth rates and endpoint diffusivity from a spatial process from which only summary statistics are available. We conclude the perspective with a discussion on the balance between under- and over-fitting, and provide a series of practical guidelines.

\section{Methods}

	\subsection{Mathematical models}

	\subsubsection{Temporal population density model}

	First, we consider a scalar model that characterises a changing population density. Denoting the population density $u(t)$, the dynamics are given by a generalised logistic equation \cite{Tsoularis.2002,Sarapata.2014},
		\begin{equation}\label{ode}
			\dv{u}{t} = r u f\!\left(\hat{u}\right),\qquad u(0) = u_0,
		\end{equation}
    where we denote $\hat{u} = u / K$ where $K$ is the carrying capacity. The \textit{crowding function}, $f(\hat{u}) \ge 0$ satisfies $f(0) = 1$, $f(1) = 0$, and $f'(\hat{u}) < \infty$ for $0 \leq \hat{u} \leq 1$. Following this choice, we can interpret $r$ as the low-density per-capita growth rate 
		\begin{equation*}
			r = \lim_{u \rightarrow 0} \dfrac{1}{u} \dv{u}{t},
		\end{equation*}
	and $K = \lim_{t \rightarrow \infty} u(t)$ as the maximum density the population can support. We work with both the low-density growth rate and crowding function, rather than directly with a density-dependent growth rate $\lambda(u) := r f(u)$, to allow more straightforward comparison between existing models, where an identical prior can be placed directly on the key quantity of interest, $r$, in all models.
			
	We make the standard assumption that observations from the population density model, denoted $u^\text{obs}(t)$, are subject to independent additive Gaussian noise with standard deviation $\sigma$ \cite{Pawitan.2001} such that
		\begin{equation*}
			u^\text{obs}(t) = u(t) + \sigma \varepsilon,
		\end{equation*}
	where $\varepsilon \sim \mathcal{N}(0,1)$ is a standard normal random variable. Similar results are expected when considering other forms of noise \cite{Murphy.2024}.

	For general $f(\cdot)$, we solve the resultant ordinary differential equation (ODE) numerically using the \texttt{DifferentialEquations} package in Julia \cite{Rackauckas.2017}. In the case that $f(\cdot)$ is piecewise linear (for example, where $f(\cdot)$ is described by a discretised Gaussian process), we exploit the analytical solution to \cref{ode} (supplementary material). 

	\subsubsection{Spatio-temporal population density model}
	
	We also consider an experiment with spatial structure, such that cell growth is captured by the generalised logistic equation and cell motility is captured through diffusion \cite{Maini.2004}. The spatio-temporal model is given by the partial differential equation	
		\begin{equation}\label{pde}
			\pdv{u}{t} = D\,\hat{\mathcal{D}}(t) \pdv[2]{u}{x} + ru\, f\left(\hat{u}\right),	
		\end{equation}
	subject to Neumann boundary conditions on the finite domain $x \in [0,L]$. Setting $f(\hat{u}) = 1 - \hat{u}$ and $\mathcal{D}(t) = 1$ recovers the well-known Fisher-Kolmogorov equation. We assume that $\hat{\mathcal{D}}(t) > 0$ and that $\hat{\mathcal{D}}(t_\mathrm{max}) = 1$ such that the model parameter $D$ corresponds to the diffusivity at the temporal end point, $t_\mathrm{max}$. 
	
	We consider two initial conditions:
	\begin{itemize}
		\item \textit{A scratch.} Representative of a scratch assay, cells are removed from the central region of an otherwise uniform monolayer. The initial condition is given parametrically by  
			\begin{equation*}
				u(x,0) = \left\{\begin{array}{ll}
					0 	& \alpha_1 L < x < \alpha_2 L,\\
					u_0 & \text{otherwise.}
				\end{array}\right.
			\end{equation*}
		The scratch location is characterised by $(\alpha_{1},\alpha_{2})$ and is assumed to be known (i.e., determined by the experimental geometry), but the initial monolayer density $u_0$ is treated as an unknown parameter.
		
		When working with the scratch initial condition, we assume that only the overall cell density, given by
			\begin{equation}\label{overall_density}
				U(t) = \dfrac{1}{L}\int_0^L u(x,t) \  \dd x,
			\end{equation}
		is reported, and is subject to additive Gaussian noise with standard deviation $\sigma_1$ such that
			\begin{equation*}
				U^\text{obs}(t) = U(t) + \sigma_1 \varepsilon.	
			\end{equation*}

		\item \textit{A step.} Representative of a moving tissue front, the initial condition is given by
			\begin{equation}\label{stepic}
				u(x,0) = \left\{\begin{array}{ll}
					u_0 	& x < 0.1L,\\
					0   & x \ge 0.1 L.
				\end{array}\right.
			\end{equation}
		When working with the step initial condition, we assume that two summary statistics are reported: the overall cell density (\cref{overall_density}), subject to additive Gaussian noise with standard deviation $\sigma_1$; and the front location, given by
			\begin{equation*}
				F(t) = \{\min x : u(x,t) < 10^{-4}\}.
			\end{equation*}		
		The front location is also subject to additive Gaussian noise, with standard deviation $\sigma_2$ such that
			\begin{equation*}
				F^\text{obs}(t) = F(t) + \sigma_2 \varepsilon.	
			\end{equation*}

	\end{itemize}
	
	For both initial conditions, we solve the time-dependent partial differential equation numerically using a central-finite-difference scheme over a uniformly spaced grid. The temporal integration is performed using the \texttt{DifferentialEquations} package in Julia \cite{Rackauckas.2017}. 
	
	\subsection{Statistical methods}

	\subsubsection{Bayesian inference approach}

	We take a standard Bayesian approach to parameter inference, assuming first that knowledge about model parameters is initially encoded in a \textit{prior} distribution, denoted $p(\bm\theta)$. We take $\bm\theta = [\varphi_1,\dots,\varphi_n,h_1,\dots,h_m]^\intercal$, to incorporate both non-negative model parameters $\varphi_i$ (i.e., $r$, $K$, $u_0$, $\beta$, $\sigma_i$), and parameters that discretise a piecewise linear function $h(x)$ such that $h(x_i) = h_i$. We place independent uniform priors on the logarithm of each model parameter, $\varphi_{i}$, to cover several orders of magnitude. We construct an appropriate prior on $\mathbf{h} = [h_1,\dots,h_m]^\intercal$ by discretising an appropriate Gaussian process prior on the function $h(x)$ (see \cref{GPmethods}).
	
	Knowledge of the parameters $\bm\theta$ is then updated using the likelihood, $\mathcal{L}$, to form the posterior distribution
		\begin{equation*}
			p(\bm\theta | \mathcal{D}) \propto \mathcal{L}(\mathcal{D} | \bm\theta) p(\bm\theta).	
		\end{equation*}
	Here, $\mathcal{D}$ denotes the data: for example, $\mathcal{D} = \{t_{i},u^\text{obs}_{i}\}$ where $t_{i}$ denotes the time and $u^\text{obs}_{i}$ the value of observation $i$. In this case, the likelihood is given by
		\begin{equation*}
			\mathcal{L}(\mathcal{D}|\bm\theta) = \prod_{i=1}^I\phi(u^\text{obs}_{i};u(t_i),\sigma^2),
		\end{equation*}
	where $I$ denotes the total number of observations, and $\phi(\cdot;\mu,\sigma^2)$ denotes the probability density function for the Gaussian distribution $\mathcal{N}(\mu,\sigma^2)$. The likelihood in cases where overall cell density and cell front measurements are made (i.e., denoted by $U_{n,i}$ and $F_{n,i}$) is constructed similarly. 
	
	We draw samples from the posterior distribution using an adaptive scaling within an adaptive Metropolis Markov chain Monte Carlo (MCMC) scheme implemented in the \texttt{AdaptiveMCMC} Julia package \cite{Vihola.2020}. Convergence of the MCMC algorithm is supported by ensuring that the improved Gelman-Rubin convergence diagnostic statistic, $\hat{R}$, satisfies $\hat{R} < 1.05$ \cite{Vehtari.2021}. To visualise model fit, we use the \textit{maximum a posteriori} (MAP) estimate, corresponding to the posterior sample with the highest posterior density. As we are primarily interested in assessing the identifiability of model parameters, all chains are initiated (where possible) using the true parameter values. In cases where the crowding function $f(\hat{u})$ is unknown even in the synthetic case, chains are initiated with the discretised logistic function $f(\hat{u}) = 1 - \hat{u}$. 
	
		\subsubsection{Gaussian process prior for unknown functions}\label{GPmethods}
		
		A distribution of a Gaussian process is, essentially, a distribution over continuous functions. We characterise the distribution of a function $h(x)$ by considering that every pair of random variables associated with function points (for example, $h(x_i)$ and $h(x_j)$) are jointly Gaussian \cite{Rasmussen.2005}. The mean of $h(x_i)$ is typically assumed to be a smooth, continuous function of $x_i$, while a \textit{kernel} describes the covariance of pairs $(h(x_i),h(x_j))$ and, in effect, determines how smooth the space of functions is. In this way, the definition of this joint distribution is independent of any particular discretisation. Furthermore,  as we demonstrate in the supplementary material (Fig. S1), the restriction that the inferred functions are sufficiently smooth leads to inferred functions that are biologically realistic and avoids overfitting.  As we detail in this section, in this work we relax the requirement that the joint distributions are unbounded Gaussian to develop priors for the non-negative crowding function $f(\hat{u})$, denoted $\mathcal{GP}_1$ and the time-dependent diffusivity $\hat{\mathcal{D}}(t)$, denoted $\mathcal{GP}_2$.

		\begin{figure}[!t]
			\centering
			\includegraphics{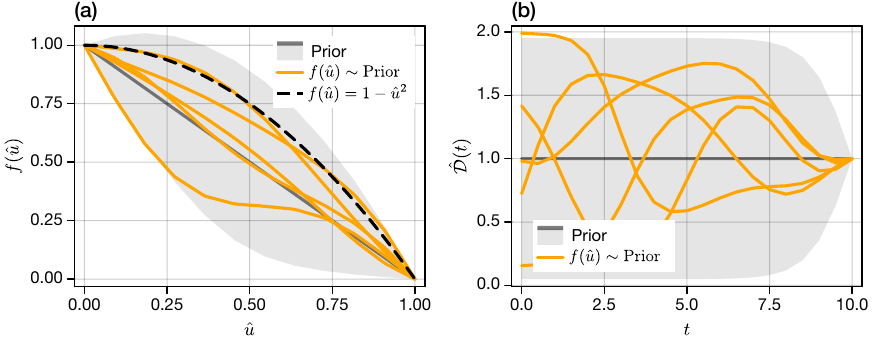}
			\caption[Figure 2]{Samples from Gaussian process priors for unknown functions $f(\hat{u})$ and $\hat{\mathcal{D}}(t)$. We apply a Gaussian process with (a) marginal truncated normal distributions and (b) marginal uniform distributions. In both cases, we show five samples from the prior (yellow), the prior mean (grey solid) and the 95\% probability interval (grey). In (a), we also show the crowding function used to generate synthetic data in \cref{fig1}, $f(\hat{u}) = 1 - \hat{u}^2$ (black dashed).}
			\label{fig2}
		\end{figure}
		
		\begin{itemize}[listparindent=\parindent]
	
		\item \textit{Crowding function.} We assume that 
				\begin{equation*}
					f(\hat{u}_i) \sim \mathcal{N}_{(0,\infty)}(\mu(\hat{u}_i), k(\hat{u}_i,\hat{u}_i))
				\end{equation*}
			where $\mathcal{N}_{(0,\infty)}$ denotes a truncated Gaussian distribution. We define the covariance structure in terms of the untruncated crowding function in which $\mathrm{cov}(\hat{u}_i,\hat{u}_j) = k(\hat{u}_i,\hat{u}_i)$. We choose
				\begin{equation*}
					\mu(\hat{u}_i) = 1 - \hat{u}_i,
				\end{equation*}
			such that the average (non-truncated) crowding function is logistic. We base the kernel on the canonical squared exponential covariance kernel \cite{Rasmussen.2005}, modified such that $\mathrm{var}(f(\hat{u}_i)) = 0$ for $\hat{u}_i \in \{0,1\}$. We have that
				\begin{equation*}
					k(\hat{u}_i,\hat{u}_j) = 2^4 \eta^2 \exp\left(-\dfrac{(\hat{u}_i - \hat{u}_j)^2}{2\rho^2}\right) \hat{u}_i\hat{u}_j(1 - \hat{u}_i)(1 - \hat{u}_j),
				\end{equation*}
			corresponding to a maximal covariance at $k(0.5,0.5) = \eta^2$. In the limit as $\rho \rightarrow 0$, we recover a prior where discretised points are independent. For demonstrative purposes, we set $\eta = 0.2$ and $\rho = 0.5$, although these choices can be inferred or constructed from \textit{a priori} knowledge about the crowding function through, for example, past experimentation \cite{Browning.2019a}.
			
			In practice, we work with a discretised piecewise-linear crowding function, such that $[f_1,\cdots,f_m]$ are inferred ($f_0 = 1$ and $f_{m+1} = 0$, by definition) where $f(\hat{u}_i) = f_i$ and $\hat{u}_i = i / (m + 1)$. In \cref{fig2}a, we compare discretised realisations of $f(\hat{u}) \sim \mathcal{GP}_1$ to the crowding function $f(\hat{u}) = 1 - \hat{u}^2$ used to generate synthetic data in \cref{fig1}, and to the prior. We take $m = 10$, which is sufficient given the prior (through the choice of scale parameter $\rho$) on the smoothness of $f(\hat{u})$ (\cref{fig2}). In the supplementary material, we demonstrate that this choice produces results very similar to those from a coarser grid with $m = 5$ (Fig. S1).

		\item \textit{Time-dependent diffusivity.} We begin by constructing an autonomous Gaussian process for $g(t)$, such that
				\begin{equation*}
					g(t_i) \sim \mathrm{Uniform}(0,1).
				\end{equation*}
			We then construct a Gaussian process for $\hat{\mathcal{D}}(t)$ as the conditional distribution
				\begin{equation*}
					\hat{\mathcal{D}}(t) = 2g(t)\, | \, 2g(t_\mathrm{max}) = 1,	
				\end{equation*}
			which ensures that $\hat{\mathcal{D}}(t_\mathrm{max}) = 1$, and that $\hat{\mathcal{D}}(t) \in (0,2)$.
			
			We construct a Gaussian process with uniform marginals by employing a Gaussian copula \cite{Nelsen.2006}. Define the Gaussian process $h(t)$ with marginals $h(t_i) \sim \mathcal{N}(0,1)$ and a standard squared-exponential kernel \cite{Rasmussen.2005}
				\begin{equation*}
					\mathrm{cov}(h(t_i),h(t_j)) = \mathrm{exp}\left(-\dfrac{(t_1 - t_2)^2}{2\rho^2}\right).
				\end{equation*}

			The Gaussian process $g(t)$ is then defined through the transformation
				\begin{equation*}
					g(t) = \phi^{-1}(h(t)),
				\end{equation*}
			where $\phi^{-1}(\cdot)$ is the quantile function (i.e., inverse distribution function) for the standard normal distribution. Similar transformation approaches could, in principle, be applied to enforce that $g(t)$ is additionally monotonic should a practitioner wish to enforce this belief; for example, by specifying $g(t)$ as the integral of the positive function $h(t)$ or by truncating the resultant distribution. However, less standard transformations are likely to lead to a less interpretable prior distribution.	
			
			In practice, we again work with a discretised piecewise-linear crowding function, such that $[g_0,\cdots,g_{m}]$ are inferred ($g_{m+1} = 0.5$, by definition) where $g(t_i) = g_i$ and $t_i = t_\mathrm{max} i / (m + 1)$. In \cref{fig2}b, we compare realisations of $\hat{\mathcal{D}}(t) \sim \mathcal{GP}_2$ to the prior, for smoothness parameter $\rho = 2$. We take $m = 19$, which is adequate given the prior (through the choice of $\rho$) on the smoothness of $\hat{\mathcal{D}}(t)$.
 
		\end{itemize}

\section{Results}

	\subsection{Misspecification with a ground truth model}

	We first expand on the example presented in \cref{fig1} by studying a population subject to generalised logistic growth with ground-truth crowding function
		\begin{equation}
			f(\hat{u}) = 1 - \hat{u}^2,
		\end{equation}
	and dynamics characterised by \cref{ode}. We independently explore the identifiability of the low-density growth rate parameter $r$ for two initial conditions: $u_0 = K / 10$, and $u_0 = K / 2$. At minimum, the parameter vector $\bm\theta = [r,K,u_0,\sigma]$ is inferred using Bayesian inference. We attempt to infer the parameters using four candidate models:
		\begin{enumerate}[label=(\arabic*)]
			\item a misspecified logistic model, such that $f(\hat{u}) = 1 - \hat{u}$;
			\item the ground truth model, with known crowding function $f(\hat{u}) = 1 - \hat{u}^2$;
			\item Richards model, $f(\hat{u}) = 1 - \hat{u}^\beta$ which can capture the ground truth through inference of an additional parameter $\beta$,
			\item and, the generalised model, where a discretised crowding function is inferred following a Gaussian process prior.
		\end{enumerate}
	Models (2)--(4) are able to capture the true underlying dynamics (albeit, for the Gaussian process approach, only in the limit as $m \rightarrow \infty$). What distinguishes models (2)--(4) is an increasing level of nested prior knowledge placed on the crowding function $f(\hat{u})$: (2) can be recovered from models (3) and (4) through very specific choices of the parameters $\beta$ and $[f_1,\cdots,f_m]$, respectively. Similarly, (3) can be recovered, approximately, from (4), through a one-dimensional family of choices for $[f_1,\cdots,f_m]$.

	\begin{figure}[!t]
		\centering
		\includegraphics{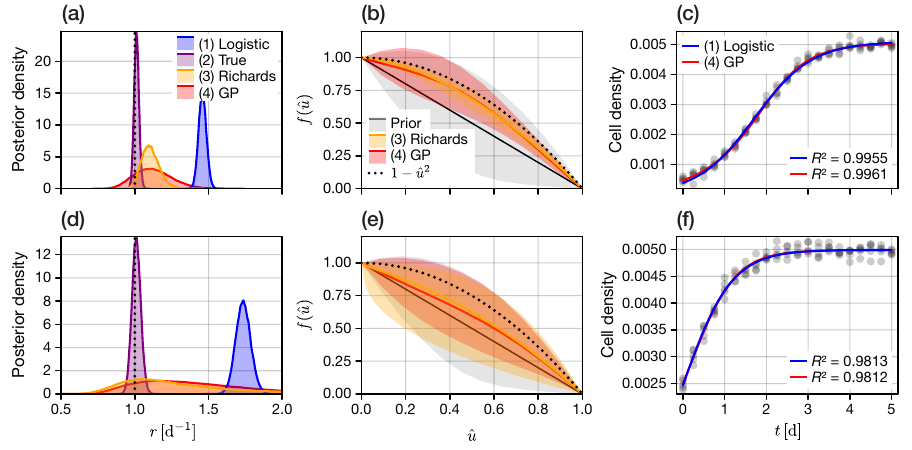}
		\caption[Figure 3]{Synthetic population growth data associated with two initial conditions is fit to four candidate models using Bayesian inference. Results in (a--c) correspond to data generated with the initial condition $u_0 = 0.5$ and in (d--f) to $u_0 = 2.5$. (a,d) Marginal posteriors for the low-density growth rate parameter, $r$, drawn from each model. The true value, $r = \SI{1}{\per\day}$ is indicated with a vertical dotted line. Results in (b,e) show the true crowding function (black dashed) $f(\hat{u}) = 1 - \hat{u}^2$ used to generate the synthetic data. Also shown are the mean and 95\% credible region for the inferred crowding function using the Gaussian process prior (red) and Richards model, where the form $f(\hat{u}) = 1 - \hat{u}^\beta$ is assumed and $\beta$ inferred (yellow). The mean and a 95\% probability interval for the Gaussian process prior are shown in grey. (c,f) Synthetic data (black semi-transparent discs, $n = 5$ replicates per time point) and the model fit at the MAP for the logistic model (blue) and the Gaussian process model (red). Goodness-of-fit is quantified using the expected value of the Bayesian $R^2$ statistic \cite{Gelman.2019}. All parameters are inferred, with true values given by $r = \SI{1}{\per\day}$, $u_0 \in \{0.5,2.5\} \,\SI{}{cells \per unit}$, $K = \SI{5e-3}{cells \per \micro\metre\squared }$, $\sigma = 10^{-4}\,\SI{}{cells \per \micro\metre\squared}$.}
		\label{fig3}
	\end{figure}

	While all parameters are inferred, in \cref{fig3} we show only the marginal posterior for the key parameter of interest, $r$, and the crowding function, $f(\hat{u})$. Most obvious from marginal posterior results from the misspecified model (1) in \cref{fig3}a,d is a clear dependence between the location of the marginal posterior and the initial condition. Not only do the posterior distributions not contain the true value, but they are non-overlapping, suggesting a statistically significant difference in low-density growth rate. Posteriors drawn for models (2)--(4) all contain the true value and, importantly, overlap between initial conditions. The strength of the corresponding assumptions that are imposed when moving from models (2) to (4) are reflected in the uncertainty in parameter estimates: estimates formed from model (2) are the most precise, while estimates drawn from the Gaussian process, which imposes the least restrictive prior assumptions, are the least precise. We note that posterior uncertainty for models (3) and (4) are, in \cref{fig3}d, similar, with support an order of magnitude larger than for models (1) and (2). This is, however, entirely expected: in \cref{fig3}d we are attempting to infer the low-density growth rate from data where cells are not observed to grow at a low density (since the initial condition is $u_0 = 2.5$). The accuracy of estimates drawn from models (1) and (2) arises primarily due to the (potentially unfounded) prior knowledge implicit to the specific choice of $f(\hat{u})$. 
	
	From \cref{fig3}a we can compare the remaining uncertainty from models (3) and (4): both models in which the crowding function is inferred. Clearly, estimates for $r$ drawn from Richards model are more precise than for the Gaussian process. The additional uncertainty that arises from a Gaussian process prior is also reflected in \cref{fig3}b, where specifying that the crowding function takes the form $f(\hat{u}) = 1 - \hat{u}^\beta$ results in a confidently and correctly inferred crowding function, as measured from the width of the credible intervals for $f(\hat{u})$ across all $\hat{u}$. In comparison, the credible intervals associated with the Gaussian process prior are a superset of those for Richards model. Notably, the posterior mean for the estimated $f(\hat{u})$ in the Gaussian process model is pulled slightly toward the logistic model through our choice of prior. Results for the less informative initial condition in \cref{fig3}e are similar: results from the Gaussian process model predominantly reflect prior knowledge, particularly for the low cell densities that are not observed.  Finally, results in \cref{fig3}c,f show that even the misspecified logistic model is almost indistinguishable from the Gaussian process model, even for the informative initial condition in \cref{fig3}c. From a single initial condition alone, model misspecification would likely remain undetected.

	\subsection{All models are wrong: misspecification without a ground truth}

	In the famous words of George Box, ``all models are wrong, but some are useful'' \cite{Box.1976}. We can expect, therefore, all mathematical models to be misspecified to some extent. To explore this in our perspective, we consider synthetic population growth data generated from a spatial process, but where only overall cell density (i.e., an average) information is available. We then explore the ability of both the necessarily misspecified logistic model, and the Gaussian process model, to estimate the low-density growth rate.	
	
\begin{figure}[!b]
		\centering
		\includegraphics{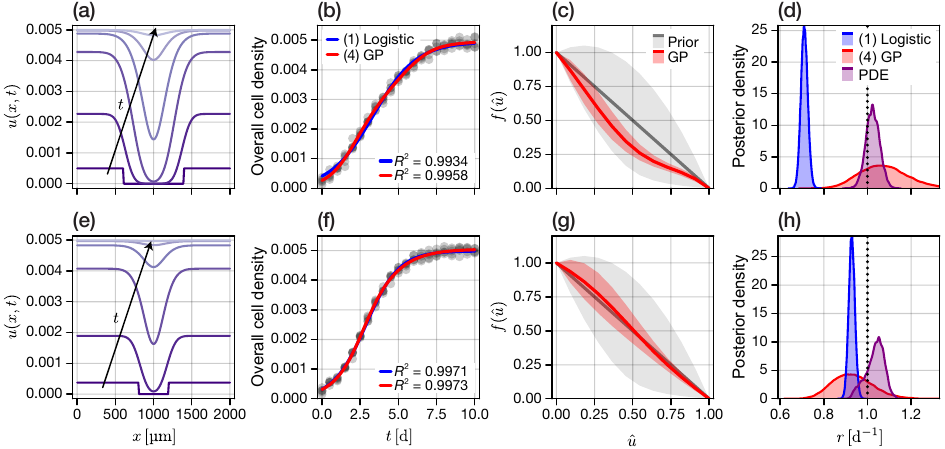}
		\caption[Figure 4]{The generalised ODE-based logistic Gaussian process model is fit to overall cell density data from a spatial process. (a,e) The underlying process is spatial, characterised by a reaction diffusion equation with constant  diffusivity and logistic growth term. (b,f) Synthetic data (black semi-transparent discs) comprises $N = 5$ measurements of overall cell density at each time point. Shown also is the model fit from the logistic ODE model at the MAP and the ODE-based generalised logistic Gaussian process model at the MAP.  Goodness-of-fit is quantified using the expected value of the Bayesian $R^2$ statistic \cite{Gelman.2019}. (c,g) The mean and 95\% credible intervals for the inferred crowding function (red) and the mean and 95\% probability intervals for Gaussian process prior (grey). (d,h) Marginal posterior distributions for $r$ from both the ODE model with Gaussian process prior (red), the PDE model (purple), and logistic model (blue), with the true value of $r$ shown as a vertical dotted line.  The parameters in the spatial model are given by $r = \SI{1}{\per\day}$, $u_0 \in \{4\times{10}^{-4},3\times{10}^{-4}\} \,\SI{}{cells \per \micro\metre\squared}$, $K = 5 \times 10^{-3}\,\SI{}{cells \per \micro\metre\squared }$, $\sigma = \SI{1e-4}{cells \per \micro\metre\squared}$. } 
		\label{fig4}
	\end{figure}

	 Specifically, we consider overall cell density data generated from synthetic \textit{scratch assay} experiments, characterised by a reaction diffusion equation with constant diffusivity and a logistic growth term (\cref{pde}). Our synthetic data set comprises two scratch assays, the only difference being the width of the scratch that is made: $(\alpha_1,\alpha_2) = (0.3,0.7)$ and $(\alpha_1,\alpha_2) = (0.4,0.6)$, as shown in Figs. \ref{fig4}a and \ref{fig4}e, respectively. We choose the initial cell density in the non-scratch region such that the overall cell densities are identical, with the denser experiment (i.e., that with an initially wider scratch) having a low initial (unknown) monolayer density of $0.1K$, where $K = \SI{5e-3}{cells \per\micro\metre\squared}$ is the (unknown) carrying capacity. A set of synthetic data ($n = 5$ replicates per time point) is shown in Figs. \ref{fig4}b and \ref{fig4}f for each initial condition.

	We first verify that the generalised logistic equation with what we term the crowding function inferred through a Gaussian process prior is able to correctly estimate the low-density growth rate. Results in \cref{fig4}d,h show that this is indeed the case, with posteriors for both initial conditions that overlap and capture the true value. Results in \cref{fig4}c,g show the crowding function inferred using data from each initial condition. We refer to this as the \textit{effective} crowding function: for spatially heterogeneous initial conditions, we do not expect to recover the true crowding function using the spatially homogeneous ODE model. Results for the second initial condition, which, with a smaller initial scratch area, is ``closer'' to the spatially homogeneous process characterised by the logistic equation, associates more closely with the prior mean of logistic growth (\cref{fig4}g). Importantly, the additional degrees of freedom in the crowding function allow the Gaussian process model to match the experimental data. It is clear from results in \cref{fig4}c that this would be difficult parametrically: not only does the crowding function depend on the (unknown) spatially heterogeneous initial condition, but it has non-monotonic derivative that would not necessarily be captured using Richards model or other parametric generalisations of the logistic growth model \cite{Tsoularis.2002}.

	In \cref{fig4}d,h, we examine the influence of additional prior knowledge on posterior uncertainty. Specifically, we compare the results from the Gaussian process model to when we assume that it is known that the process follows the reaction-diffusion equation, and that the initial scratch location, through $(\alpha_1, \alpha_2)$, is also known. While it is still the case that only overall cell density measurements are available, the posterior uncertainty is reduced significantly in comparison to that associated with the ODE-based Gaussian process model. We highlight, however, that not only is the ODE-based Gaussian process approach computationally cheaper than that of the PDE, but the assumptions made likely much more realistic. Many low density experiments, for instance, may be associated with some level of spatial heterogeneity (for instance, through proliferation, which is a local process). Not only is it likely that this structure varies between experimental replicates, but in the case that only overall cell density information is available at later time points, there is no reason to expect that information about the spatial structure (or, indeed, the spatial process itself) be known.
	
	\begin{table}[!b]
		\centering
		\caption{MAP estimates and 95\% credible intervals for the low-density growth rate parameter $r$, for data sets with $N$ observations per time point. In all cases, the true value is given by $r = \SI{1}{\per\day}$. The location of the scratch is given by $(\alpha_1,\alpha_2) = (0.3,0.7)$ and $(\alpha_1,\alpha_2) = (0.4,0.6)$ for initial conditions 1 and 2, respectively.}
	    \renewcommand{\arraystretch}{1.2}
		\label{tab1}
		\begin{tabular}{|r|cc|cc|}\hline
	    & \multicolumn{2}{c|}{\bfseries Initial Condition 1} & \multicolumn{2}{c|}{\bfseries Initial Condition 2} \\\hline\hline
	    \multicolumn{5}{|c|}{\bfseries $f(\hat{u}) = 1 - \hat{u}$}\\\hline
$N$ & MAP    & 95\% Credible Interval & 
	  MAP    & 95\% Credible Interval \\\hline
5   & 0.6727 & (0.6481,0.6985) & 0.8945 & (0.8679,0.9199) \\
50  & 0.6898 & (0.6817,0.6987) & 0.9181 & (0.9090,0.9273) \\
100 & 0.6901 & (0.6844,0.6965) & 0.9141 & (0.9075,0.9203) \\
200 & 0.6930 & (0.6885,0.6973) & 0.9134 & (0.9089,0.9180) \\
\hline\hline
	    \multicolumn{5}{|c|}{\bfseries $f(\hat{u}) \sim \mathrm{GP}$}\\\hline
$N$ & MAP    & 95\% Credible Interval & 
	  MAP    & 95\% Credible Interval \\\hline
5   & 0.9274	& (0.7719,1.1193) &	0.8777 & (0.7495,1.1650)\\
50  & 0.9210	& (0.8250,1.0226) &	1.0023 & (0.8908,1.1208)\\
100 & 0.9832	& (0.8914,1.0638) &	0.9933 & (0.8856,1.0721)\\
200 & 0.9976	& (0.9090,1.0540) &	1.0422 & (0.9515,1.1116)\\
\hline
	\end{tabular}
    
	\end{table}
	
	Finally, we demonstrate how misspecification can lead to over confidence, by gradually increasing the amount of synthetic data calibrated to both the misspecified logistic model, and to a model with a crowding function inferred through the Gaussian process prior. For our choice of likelihood, this approach is analogous to decreasing the measurement error noise. Specifically, in an additive Gaussian model, decreasing the measurement error by a factor of two, for instance, corresponds approximately to increasing the number of data points by a factor of four \cite{Pawitan.2013}.
	
	Results in \cref{tab1} show that estimates for the low-density growth rate drawn from the logistic model vary between initial conditions, and converge to incorrect values as the number of data points is increased. The level of confidence in the (incorrectly) inferred growth rates is concerning: for both initial conditions, the 95\% credible interval width is less than 1\% of the inferred value. Statistically, one could erroneously, yet reasonably, conclude a physiological difference in the low-density growth between these experiments, despite the only difference being due to the initial condition. 	Estimates obtained using a Gaussian process prior for the crowding function, however, are more positive: the credible intervals all contain the true value, even for very large data sets ($N = 200$ per time point). Most importantly, the remaining non-identifiability in the crowding function results in credible intervals that do not appear to converge as rapidly. Even for the largest data set, the credible interval widths remain at approximately 10\% of the inferred value. This is representative of remaining model uncertainty. For the spatial process, it is straightforward to see why we might expect this: even an infinite number of overall cell density measurements cannot be expected to yield information about spatial structure. The question of \textit{structural identifiability} in spatial problems subject to scalar measurements remains an open problem \cite{Browning.2024b}.

	\subsection{Time-dependent parameters}	
	The final scenario we explore is where misspecification arises due to parameters that depend upon time. In cell biology, for example, the start of an \textit{in vitro} experiment is often associated with a disturbance, with a time-delay or lag-phase until cell behaviour normalises \cite{Jin.2017}. For example, in a cell proliferation assay the cells may require some time to adapt to the glass substrate before dividing. Similarly, in a scratch assay, cells may require some time to adapt before migrating and dividing. Existing parametric approaches typically assume that such time-dependence can be captured by scaling the diffusivity and/or proliferation terms in a reaction diffusion equation with time-dependent parametric functions, for example $\tanh(\alpha t)$ in \cite{Simpson.2024b}, or $1/\exp(-\alpha_{1} - \alpha_{2}t)$  in \cite{VandenHeuvel.2022}, where the $\alpha_{i}$ characterise the delay. Alternative parametric approaches that are focused on ODEs assume that the dynamics are biphasic, with model parameters that transition frdsom an initial to a final value at a switch point that is estimated \cite{Murphy.2022a}. Lagergren et al. \cite{Lagergren.2020} approach the reaction diffusion equation problem non-parametrically, using a BINN approach to machine learning to learn a time-dependent function that scales both the diffusivity and proliferation rate terms. 

\begin{figure}[!t]
		\centering
		\includegraphics{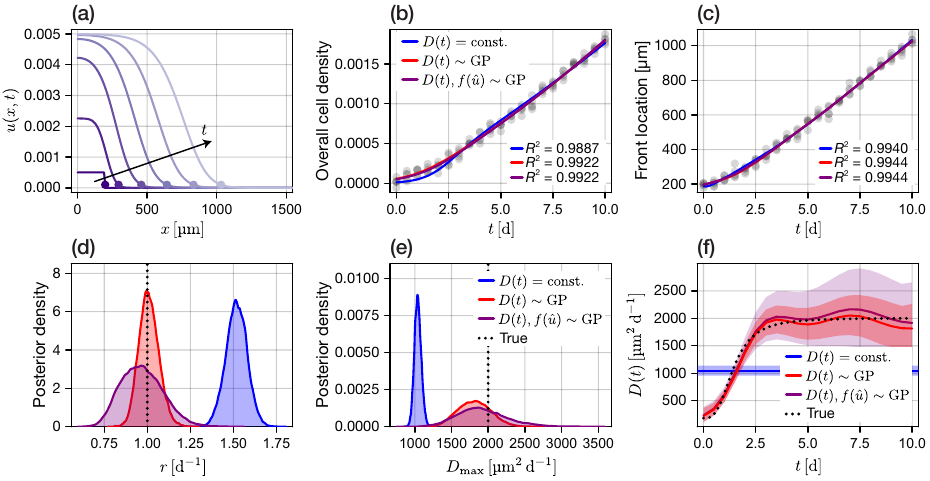}
		\caption[Figure 5]{Calibration to data with a disturbance phase, corresponding to a time-dependent parameter. (a) We consider a spatial process governed by a reaction-diffusion equation with logistic growth and a time-dependent diffusivity. Shown also are the corresponding cell front locations (discs). (b--c) Noisy synthetic measurements of (b) the overall cell density and (c) the front location (grey discs). Shown also is the model fit from the constant diffusivity (blue), a model with time-dependent diffusivity (red) at the MAP. Goodness-of-fit is quantified using the expected value of the Bayesian $R^2$ statistic \cite{Gelman.2019}. (d--e) Marginal posterior distributions from each model, along with the true parameter values (vertical dotted line).  Shown also are results from a model where both a time-dependent diffusivity and density-dependent crowding function are inferred (purple). (f) The inferred constant diffusivity is shown in blue, along with a time-dependent diffusivity inferred assuming a known crowding function (red) and in a model where the crowding function is also inferred (purple).  In each case, we show the mean and a 95\% credible interval. The true diffusivity (\cref{Dhat}) is also shown (black dotted). }
		\label{fig5}
	\end{figure}

	We consider a spatial process governed by a reaction-diffusion equation (\cref{pde}) subject to logistic growth and a time-dependent diffusivity that captures an initial disturbance phase. The diffusivity is given by $D(t) = D_\mathrm{max}\hat{\mathcal{D}}(t)$ where
		\begin{equation}\label{Dhat}
			\hat{\mathcal{D}}(t) = \dfrac{k(t)}{k(10)},\qquad k(t) = \dfrac{t^3}{t^3 + 3} + 0.1,
		\end{equation}
	such that the diffusivity at the start of the experiment is approximately 10\% of the maximum when the experiment concludes at $t = \SI{10}{\day}$ (\cref{fig5}g). The transition from the initial to the final diffusivity is smooth, and occurs at approximately $t \sim 3^{1 / 3}\,\SI{}{\day}$. The functional form in \cref{Dhat} is chosen such that it would be difficult to \textit{a priori} propose a suitable parametric form. At $t=0$, the cell density is given by a step function (\cref{stepic}) with an observed and clearly defined front, although the initial cell density behind the front is unknown. Two measurements are available across multiple time points: the overall cell density, and the location of the cell front. A spatially resolved simulation, and a set of synthetic data, are shown in Figs. \ref{fig5}a to \ref{fig5}c.

	We explore the effect of misspecification on the estimation of the low-density growth rate, $r$, and the maximum diffusivity, $D_\mathrm{max}$, by calibrating both a misspecified model with a constant diffusivity \cite{Maini.2004}, and a model where the time-dependent diffusivity is inferred through a Gaussian process prior on $\hat{\mathcal{D}}(t)$. Results in Figs. \ref{fig5}b and \ref{fig5}c, show only subtle differences between the ability of each model to recapture overall cell density and front location data. While examination of the residuals would likely suggest model misspecification, it could easily go unnoticed without. The marginal posterior for the parameters of interest in Figs. \ref{fig5}e and \ref{fig5}f show that only the time-dependent model is able to capture the true value of each parameter. Furthermore, the Gaussian process approach is able to, with some remaining uncertainty, estimate the time-dependent diffusivity (\cref{fig5}g). The result from calibration to the misspecified (i.e., constant-diffusivity) model again show over confidence around an incorrect value. While the misspecified model provides a potentially adequate fit to this data, it is unlikely to extrapolate outside the range of the data to produce reliable predictions. We might also expect, for example, estimates to vary if data were stratified to include or exclude the first $t \approx \SI{3}{\day}$ of the experiment during the disturbance phase. 

Finally, we explore a scenario where the crowding function $f(\hat{u})$ is also unknown, and must be inferred alongside the time-dependent diffusivity. Results in \cref{fig5}d,e demonstrate that the reduction in prior knowledge is associated with increased uncertainty in parameter estimates. The time-dependent diffusivity, shown in \cref{fig5}f is also inferred, now with greater uncertainty.

\section{Discussion and conclusions}

Complex biological systems necessitate correspondingly complex models. Available data, however, are often limited: even relatively simple mathematical models can be practically non-identifiable. A persistent challenge in mathematical biology is to balance model complexity and identifiability without exacerbating issues related to model misspecification. Complex or over-parameterised models may provide superior fits, but the lack of parameter identifiability limits biological insight. Simpler misspecified models, on the other hand, may lead to precise but biased parameter estimates and potentially misleading conclusions. This phenomenon is analogous to Simpson's paradox \cite{Simpson.1951} in statistics: incorporating additional freedom in a misspecified model can, potentially substantially, alter estimates of other parameters. The primary goal, therefore, should not be the \textit{precise} estimation of model parameters, but the \textit{accurate quantification of the uncertainty} in parameter estimates.

While model misspecification can be detected through validation (i.e., prediction and comparison using unseen data), or through a residual analysis \cite{Simpson.2022,Lambert.2023}, it must still be resolved. We take a semi-parametric approach to resolve potential misspecification, using a Gaussian process prior to delineate parameters of interest from the unknown functional form of the model itself. In the context of misspecification, this approach allows us to quantify prior knowledge about model terms by specifying only their \textit{smoothness}: we view this as a continuous and non-parametric analogue of traditional model selection or multimodel techniques \cite{Linden-Santangeli.2024}. Importantly, formulating the model in this way demonstrates how prior knowledge explicitly or implicitly always plays a role. Even specifying a parametric form for a term (for example, Richards model assumes a crowding function of the form $1 - x^\beta$) is a strong assumption, comparable to fixing an unknown model parameter to a particular value. In practice, priors for the logistic crowding function or the time-dependent delay in our latter example could be constructed from existing posteriors, thereby explicitly incorporating existing knowledge \cite{Browning.2019a}.

Most obvious in our results is that parameter estimates drawn from a Gaussian process model are associated with more uncertainty. For population growth rate data, prespecifying the exact form of the crowding function can result in very precise parameter estimates (see \cref{fig3}), albeit at great risk: estimates drawn from an incorrectly specified crowding function can be biased, with clear dependence upon the initial condition (see \cref{fig1,fig3}). Furthermore, specifying only the correct parametric form (i.e., through Richards model) results in relatively precise parameter estimates, however is a potentially unfounded assumption. Estimates drawn from the Gaussian process model using only high-density data are imprecise, with a posterior for the low-density growth rate varying across several orders of magnitude (\cref{fig3}d). This, however, is entirely expected: outside of prior knowledge, there is no reason for high-density data to contain information about low-density behaviour. The artifactual identifiability of the low-density growth rate from all models except the Gaussian process model in \cref{fig3}d is attributable to model choice and the result of prior knowledge. Our results consistently demonstrate that any reduction in prior knowledge results in more uncertain parameter estimates. Even in \cref{fig5}, where we infer \textit{both} the density-dependent diffusivity and crowding function, we have enforced that the model dynamics follow a reaction-diffusion equation. 

To some extent, we should expect that all models are necessarily misspecified \cite{Box.1976,Enderling.2021}. We explore this by applying ODE-based models to population density data arising from a spatially heterogeneous process: a pair of scratch assays. Application of the classical but misspecified logistic model (which forms the ground truth in the spatially homogeneous case) leads to the conclusion that cell behaviour differs between the two experiments. However, our approach with a Gaussian process allows the ODE model to capture an effective crowding function that differs between experiments and provides more robust and accurate parameter estimates that remain identifiable. In the limit that a large data set is considered, posteriors for the misspecified logistic model notably narrow very tightly around biased parameter estimates. This is not the case for the Gaussian process model: while posteriors always contain the true value, they do not narrow as rapidly around any value. We attribute this residual uncertainty, even in the limit that the cell density is (effectively) observed precisely, to the fact that there remains no information about the spatial structure. In practice, it is unlikely that biologically processes can be observed completely. We should not expect, therefore, for uncertainty in parameter estimates to vanish in the large data limit.

The Gaussian process approach provides a natural and interpretable extension to standard Bayesian inference methods. Key to their success in dealing with potential misspecification is that they provide flexibility that allows the model to capture the underlying dynamics. Alternative non-parametric methods that also allow for this are primarily based on neural networks. In particular, Lagergren et al. \cite{Lagergren.2020} demonstrate how BINNs can be applied to infer density-dependent diffusivities and growth rates in a reaction diffusion equation. Outside of deterministic models, Jo et al. \cite{Jo.2024} apply a neural network approach to infer the delay distribution within a stochastic model of cellular signalling. In both studies, the authors alleviate model misspecification by non-parametrically inferring unknown functions of distributions. Neural network approaches do not require the user to (explicitly) specify a prior over a space of functions. This benefit, however, comes with a reduced ability to quantify inferential uncertainty. New approaches, based on Bayesian neural ODEs promise to address this problem, however, as with the Gaussian processes, are computationally expensive \cite{Dandekar.2022}.

Misspecification can also be accounted for indirectly through more sophisticated statistical methods; for example, by capturing the underlying correlation structure in the residuals \cite{Lambert.2023,Amadi.2023}. Approximate Bayesian computation methods \cite{Wilkinson.2013,Thomas.2025} (in addition to those based on a neural likelihood \cite{Kelly.2024}, and control theory \cite{Clairon.2021}) are also able to account for model misspecification, by relaxing the typical requirement that the model must fit in the infinite data limit. More explicitly, approaches based on \textit{gradient matching} incorporate (typically additive) terms that capture model misspecification to the ODE right-hand-side \cite{Ellner.2002,Dondelinger.2013}; a Gaussian process is then applied to fit ODE models in state space, without the requirement that the ODE is followed exactly \cite{Hooker.2015,Wenk.2020}. These statistical approaches typically provide wider confidence intervals and prevent over-confidence in biased parameter estimates. In contrast, by delineating parameter of interest from misspecification, we are, in some cases, able to draw relatively precise estimates from an otherwise misspecified model: for a spatial process with temporally varying diffusivity, posteriors for the low-density growth rate drawn from both a Gaussian process and misspecified model have similar variance.

A common reality in mathematical biology is that systems are complex and data are limited. Important model calibration goals should include accurate quantification of uncertainty in both parameter estimates and the underlying model. Accurately quantifying the uncertainty in parameter estimates is essential for both obtaining biological insight and designing future experiments \cite{Faller.2003,Jang.2025}. As we demonstrate in this perspective, it is particularly important to formulate priors that better represent existing knowledge. While simplification can often lead to identifiability, resultant misspecification can give rise to misleading insights. Fortunately, identifiability is not necessarily required for accurate prediction \cite{Simpson.2024a}, nor is identifiability of all model parameters requisite for biological insight.

\section*{Acknowledgements}

J.A.F. acknowledges support from the Australian Research
Council (FT210100034, CE230100001).

\section*{Data availability}

Code used to produce the results is available at \url{https://github.com/ap-browning/misspecification}.

		
\end{document}


\clearpage
\section*{\centering Supplementary Material}

\renewcommand{\theequation}{S\arabic{equation}}
\renewcommand{\thesection}{S\arabic{section}}
\renewcommand{\thefigure}{S\arabic{figure}}
\renewcommand{\thetable}{S\arabic{table}}
\setcounter{section}{0}
\setcounter{figure}{0}
\setcounter{equation}{0}

\section{Analytical solution to the population growth model}

Here, we construct the analytical solution to the non-dimensional generalised logistic equation
	\begin{equation}
		\dv{\hat{u}}{t} = \hat{u} g(\hat{u}),\qquad \hat{u}(0) = \hat{u}_0.
	\end{equation}
in the case that $g(\hat{u})$ is piecewise linear on $\hat{u} \in (\hat{u}_0,1)$ such that $g(\hat{u}_i) = g_i > 0$ at a set of discrete ordered points $\hat{u}_0 < \hat{u}_1 < \cdots < 1$, and that $g(1) = 0$. We therefore have that
	\begin{equation}
		g(\hat{u}) = a_i \hat{u} + b_i, \qquad \hat{u}_i \le \hat{u} < \hat{u}_{i+1},
	\end{equation}
with
	\begin{equation}
		a_i = \dfrac{g_{i+1} - g_i}{\hat{u}_{i+1} - \hat{u}_i}, \qquad b_i = g_i - a_i \hat{u}_i.
	\end{equation}

Consider that, for $t \in [\tau_i, \tau_{i+1}]$,
	\begin{align*}
		\dv{\hat{u}}{t} &= a_i \hat{u}^2 + b_i \hat{u}, \qquad \hat{u}(\tau_i) = \hat{u}_i.
	\end{align*}
Integrating the separable equation yields
	\begin{equation}\label{si_integrated}
		b_i(t - \tau_i) = \log\left(a_i + \dfrac{b_i}{\hat{u}_i}\right)-\log\left(a_i + \dfrac{b_i}{\hat{u}(t)}\right).
	\end{equation}

We can now recursively solve for $\tau_{i}$ by defining $\tau_i$ such that $\hat{u}(\tau_i) = \hat{u}_i$ is true for all $t$. This definition is unique by our previous assertion that $g(\hat{u})$ is strictly positive for $\hat{u}_{1} \leq \hat{u} < 1$. Therefore,
	\begin{align*}
		\tau_{i+1} &= \tau_i + \dfrac{1}{b_i}\left[\log\left(a_i + \dfrac{b_i}{\hat{u}_i}\right)-\log\left(a_i + \dfrac{b_i}{\hat{u}_{i+1}}\right)\right],
	\end{align*}
with $\tau_0 = 0$. Inverting \cref{si_integrated} and simplifying yields
	\begin{equation}
		\hat{u}(t) = \dfrac{b_i}{\mathrm{e}^{-b_i (t - \tau_i)}\left(a_i + \dfrac{b_i}{\hat{u}_i}\right) - a_i}
	\end{equation}
which applies for $t \in [\tau_i, \tau_{i+1}]$.


\clearpage
\section{Gaussian process resolution}

\begin{figure}[H]
	\centering
	\includegraphics{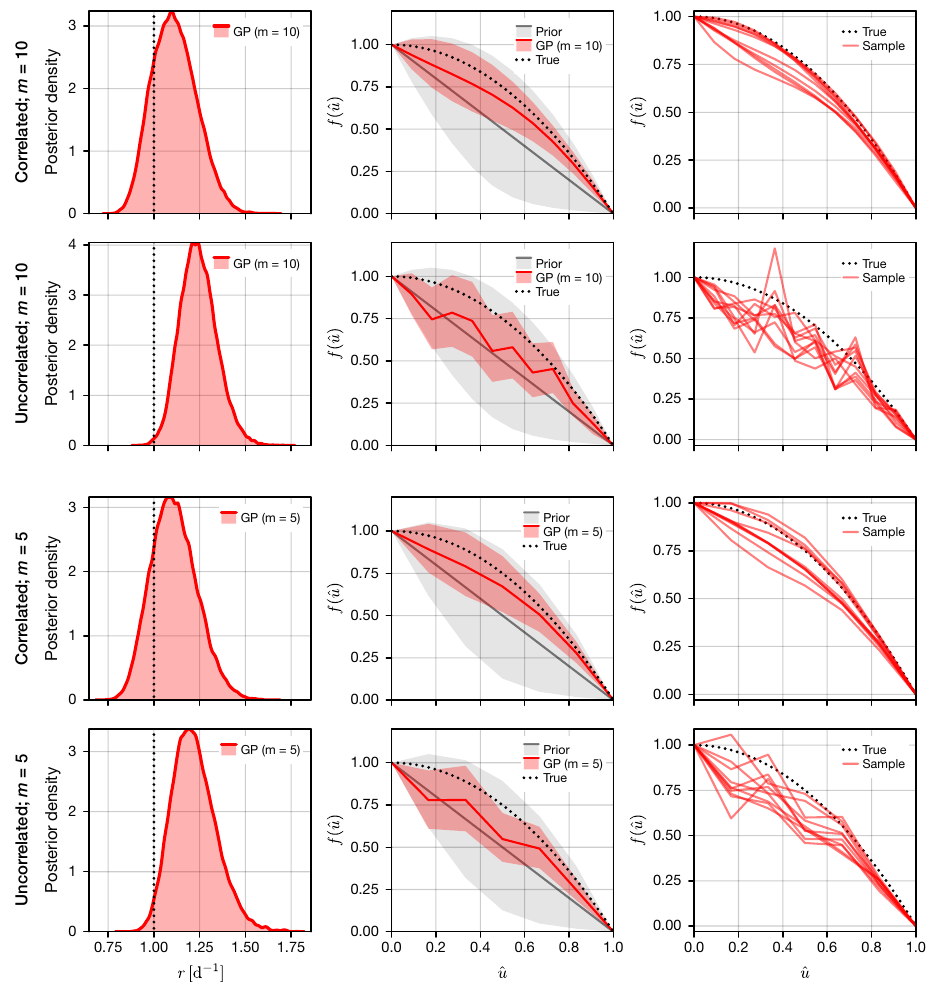}
	\caption{We reproduce the analysis of Fig. 3a,b in the main text for two discretisations of Gaussian process crowding function ($m = 5$ and $m = 10$). Results in the first row correspond to that in the main text. Additionally, we explore the case that $\rho \rightarrow 0$: a scenario in which the prior distribution is uncorrelated between discretisation points. Results show very little difference between a discretisation with $m = 10$ and $m = 5$ for the correlated prior used in the main text. Results using an uncorrelated prior do not lead to smooth crowding functions. }
	\label{figS1}	
\end{figure}